\documentclass{article}

\usepackage[preprint]{neurips_2024}

\usepackage[pdftex]{graphicx}
\usepackage{makecell}
\usepackage[dvipsnames]{xcolor}
\usepackage[utf8]{inputenc} 
\usepackage[T1]{fontenc}    
\usepackage{hyperref}       
\usepackage{url}            
\usepackage{booktabs}       
\usepackage{amsfonts}       
\usepackage{nicefrac}       
\usepackage{microtype}      
\usepackage{xcolor}         
\usepackage{float}
\usepackage{array}
\usepackage[raster,most]{tcolorbox}
\usepackage{natbib}
\usepackage{multirow}
\usepackage{wrapfig}
\usepackage{amssymb}
\usepackage{pifont}

\usepackage{cleveref}

\definecolor{xiaomiblue}{HTML}{4A7BCE}      
\definecolor{xiaomipaleblue}{HTML}{B8DCFE}  
\definecolor{xiaomiorange}{HTML}{FFA903}    
\definecolor{xiaomiteal}{HTML}{03CCA0}      
\definecolor{xiaomigreen}{HTML}{50B341}     
\definecolor{xiaomicoral}{HTML}{ED696D}     
\definecolor{xiaomilightgray}{HTML}{AAAAA8} 
\definecolor{xiaomibrightblue}{HTML}{04A3FD}
\definecolor{xiaomimedgray}{HTML}{6E6E6C}   
\definecolor{xiaomiblack}{HTML}{030303}     
\definecolor{xiaomired}{HTML}{ee4028}

\newcommand{\xmark}{\textcolor{xiaomired}{\text{\ding{55}}}} 
\newcommand{\cmark}{\textcolor{xiaomigreen}{\text{\ding{51}}}} 
\setcitestyle{numbers,square}

\title{\textcolor{xiaomiorange}{Mi}DashengLM: Efficient Audio Understanding with General Audio Captions}

\author{
MiLM Plus \\
\\
Xiaomi Inc., China\\
}

\begin{document}

\maketitle

\setcounter{footnote}{0}

\begin{abstract}
Current approaches for large audio language models (LALMs) often rely on closed data sources or proprietary models, limiting their generalization and accessibility.
This paper introduces MiDashengLM, a novel open audio-language model designed for efficient and comprehensive audio understanding through the use of general audio captions using our novel ACAVCaps training dataset. 
MiDashengLM exclusively relies on publicly available pretraining and supervised fine-tuning (SFT) datasets, ensuring full transparency and reproducibility. 
At its core, MiDashengLM integrates Dasheng, an open-source audio encoder, specifically engineered to process diverse auditory information effectively. 
Unlike previous works primarily focused on Automatic Speech Recognition (ASR) based audio-text alignment, our strategy centers on general audio captions, fusing speech, sound and music information into one textual representation, enabling a holistic textual representation of complex audio scenes.
Lastly, MiDashengLM provides an up to 4$\times$ speedup in terms of time-to-first-token (TTFT) and up to 20$\times$ higher throughput than comparable models.
Checkpoints are available at \href{https://github.com/xiaomi-research/dasheng-lm}{\includegraphics[height=1em,keepaspectratio]{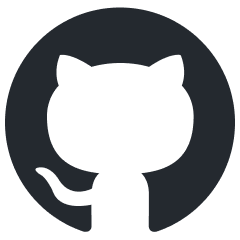}} and \href{https://huggingface.co/mispeech/midashenglm-7b}{\includegraphics[height=1em,keepaspectratio]{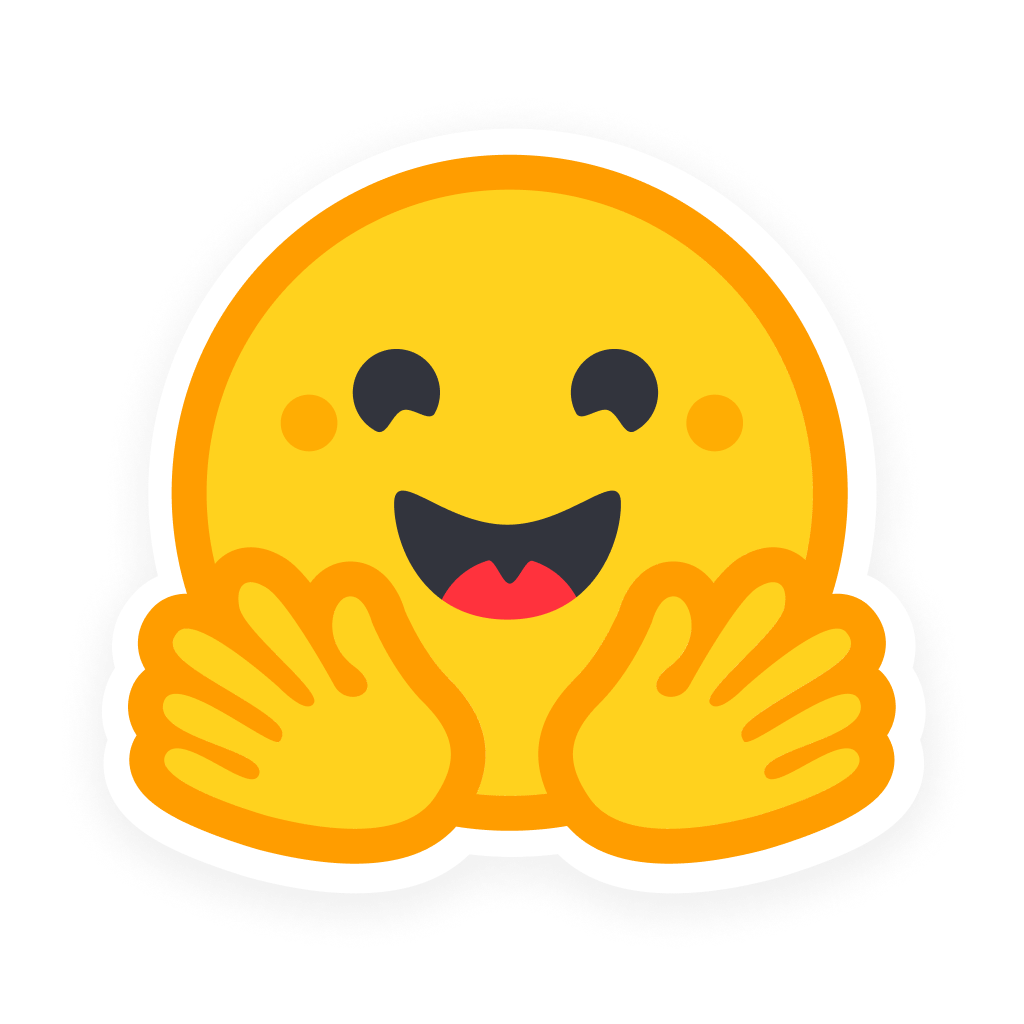}}.
\end{abstract}

\begin{figure}[H]
    \centering
    \includegraphics[width=0.77\linewidth]{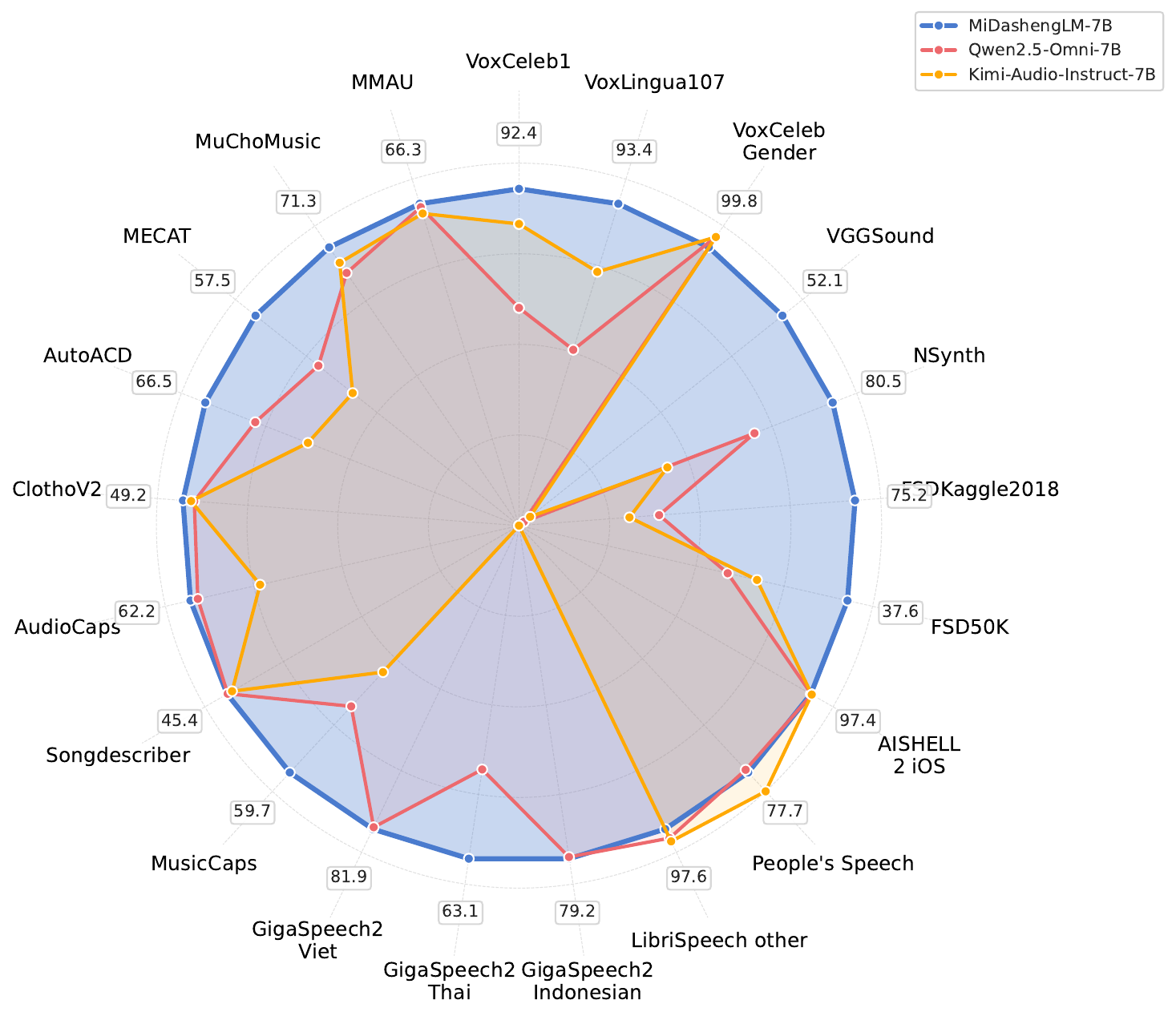}
\end{figure}

\section{Introduction}

Large language models (LLMs) have played a pivotal role in advancing machine learning approaches for natural language processing (NLP), demonstrating impressive capabilities in understanding the world through text. 
While these models can effectively interact with humans via text, the ability to understand sound remains crucial for agents to fully engage with the physical world.
Large Audio-Language Models (LALMs) aim to bridge the gap between auditory and textual understanding.
Within the audio domain, we identify three commonly used broad categories: speech, (environmental) sounds and music.
Aligning audio with text requires a mapping between speech/sound/music and respective text.
For speech the most common alignment are transcripts, while captions are used for sound and music.
Transcripts can be understood as a monotonous alignment between audio and text domains.
In contrast, captions are typically used for broader audio elements like sounds and music, offering a more generalized alignment, meaning they capture the overall nature or occurrence of a sound.

Current audio understanding research typically processes speech transcripts, audio captions, and music captions separately. 
This independent approach limits the depth and completeness of auditory scene analysis.
Anther key limitation stems from existing audio captions, which often offer only superficial descriptions. 
For example, spoken content is frequently simplified to ``somebody is speaking'', ignoring semantic details.
Furthermore, these datasets often fail to capture critical auditory aspects like room acoustics (e.g., reverberation) or signal quality. 

To overcome these limitations, this paper proposes fusing speech transcripts, audio captions, and music captions into a single, unified general caption. 
Our goal is to create a holistic textual representation that jointly includes all relevant audio information, providing a more detailed and semantically rich description of the auditory environment.

\subsection{Motivation}
\label{sec:motivation}

Developing a LALM requires aligning audio features with textual descriptions.
Utilizing sound and music captions as a training target has been previously explored~\cite{deshmukh2023pengi,kong2024audio,ghosh2024gama,ghosh2025audio} to enhance audio understanding.
However, these approaches lack automatic speech recognition (ASR) capabilities, limiting their usefulness for general applications, as users expect a LALM to handle both general audio understanding and speech — not just captions.
The most used alignment paradigm couples large language models (LLMs) with audio understanding through automatic speech recognition (ASR). 
This approach prevails for two key reasons: First, numerous high-quality off-the-shelf ASR models exist that can generate reasonably accurate transcripts automatically. 
Second, a substantial portion of internet audio content consists of speech-based material - including podcasts, lectures, interviews, and other spoken-word formats - making ASR an effective bridge between audio and text modalities.
Several prominent works have demonstrated the effectiveness of ASR-based LALM training, such as Whisper~\cite{radford2023robust}, SpeechT5~\cite{ao2021speecht5}, Universal Speech Model (USM)~\cite{zhang2023google}, Open Whisper-style Model (OWSM)~\cite{peng2025owsm} and Kimi-Audio~\cite{kimi_audio}.
However, we argue that ASR-based pretraining provides limited benefits for general audio-language understanding, due to the following reasons:

\paragraph*{Inefficient Data Utilization}
Large-scale pretraining on million-hour long datasets typically relies on existing automated speech recognition (ASR) pipelines to generate transcripts from speech. 
This results in a substantial loss of potentially valuable data, as sounds like music, environmental noises, or even silent pauses are discarded.
Using a general captioning approach has the benefit that any audio can be used for training, as even ``noisy'' audio clips could be labeled.
This significantly enhances data diversity, allowing models to learn from a much wider range of acoustic information beyond just speech.

\begin{figure}[h]
    \centering
    \includegraphics[width=0.6\linewidth]{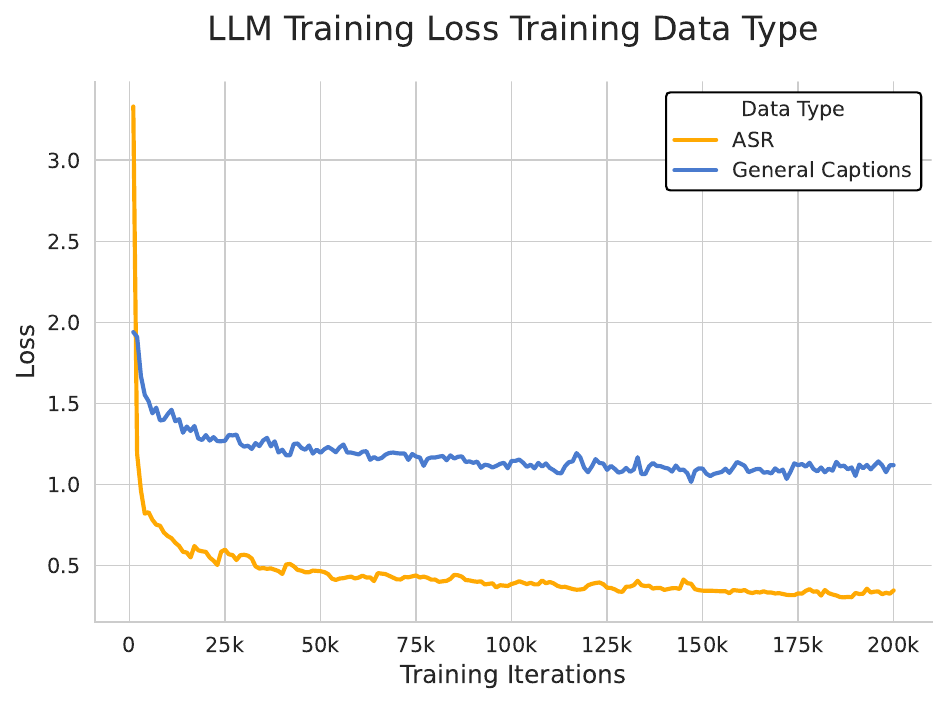}
    \caption{Training cross entropy loss (next token) curves between ASR and caption based pretraining. General captions utilize the ACAVCaps (\Cref{tab:env_data}) dataset, while ASR uses ACAV100M-Speech (\Cref{tab:speech_data}). ACAV100M-Speech contains up to 90 different languages, while captions are English only.}
    \label{fig:pretraining_losses_asr_vs_caption}
\end{figure}

\paragraph*{Trivial objective} The training losses for ASR-based LALMs are typically low, even across different languages, suggesting that the models learn relatively little meaningful information from ASR-based data, compared to text-based training~\cite{touvron2023llama} (see \Cref{fig:pretraining_losses_asr_vs_caption}).
We attribute this to the simplicity of speech-text alignments, where the temporal ordering of acoustic units and their corresponding text tokens follows a monotonic  (left-to-right) correspondence.
Thus a model only needs to establish local correspondences between spoken words and their textual counterparts, bypassing the need to understand broader (global) audio context.

\paragraph*{Limitations of ASR-Based Pretraining Beyond Speech Content} 
ASR-based pretraining does not focus on information other than the spoken content.
This limited scope means that important speech meta-information, such as a speaker's gender, age, or emotional state, is not captured or integrated during the pretraining process.
Furthermore, the pretraining methodology overlooks audio signal-specific characteristics like reverberation levels, recording quality, and environmental acoustics.

\subsection{Audio caption and speech summarization}

Audio captions have been the focus of extensive research~\cite{wu2019audio,drossos2020clotho,drossos2017automated}.
Most datasets during the start of the audio-caption era were manually labeled~\cite{audiocaps,wu2019audio,drossos2020clotho,martin2021ground_macs2}, but recent work has leveraged large language models (LLMs) to scale and streamline dataset creation.
Notable LLM-assisted audio captioning datasets include WavCaps~\cite{mei2023wavcaps}, AutoACD~\cite{sun2024autoacd}, SoundVECaps~\cite{yuan2024soundvecaps}, AudiosetCaps~\cite{bai2024audiosetcaps} and FusionAudio-1.2M~\cite{chen2025fusionaudio}.

These works utilized LLMs in order to enhance existing audio captions by additional visual information~\cite{yuan2024soundvecaps}, temporal information~\cite{sun2024autoacd} or with additional CLAP filtering~\cite{bai2024audiosetcaps,ghosh2024gama}.
However, we identify two key limitations in existing datasets:

Neglect of spoken language: Publicly available captioning data primarily focuses on sound/music events and their audio-visual/temporal relationships, despite speech constituting the majority of real-world audio~\cite{gemmeke2017audio}.
Current audio captioning datasets can therefore be better understood as (environmental-) sound captioning datasets.
Limited data diversity: Popular datasets (AudioCaps, WavCaps, AutoACD, SoundVECaps, AudiosetCaps and FusionAudio-1.2M) predominantly derive from the same audio sources (Audioset~\cite{gemmeke2017audio}, VGGSound~\cite{chen2020vggsound} and FSD50k~\cite{fonseca2021fsd50k}).
This source overlap leads to a problematic one-to-many mapping: multiple ``distinct'' datasets are, in fact, derived from identical underlying audio clips, containing different textual descriptions.
This redundancy adds little training audio data variation, limiting model generalization.

While audio captions have been used for LALM pretraining, existing approaches typically generate new captions through either (1) paraphrasing existing descriptions~\cite{ghosh2024gama,audioflamingo3} or (2) augmenting them with (unrelated) video context~\cite{ghosh2025audio} using LLMs, rather than genuinely diversifying the underlying audio content.

In our work we rely on \textit{general audio captions}, a novel captioning type.
General audio captions can be understood as a fusion of speech summarization~\cite{retkowski2025speech}, music captions and audio captions into one.

\section{Framework}

\begin{figure}[t]
    \centering
    \includegraphics[width=0.84\linewidth]{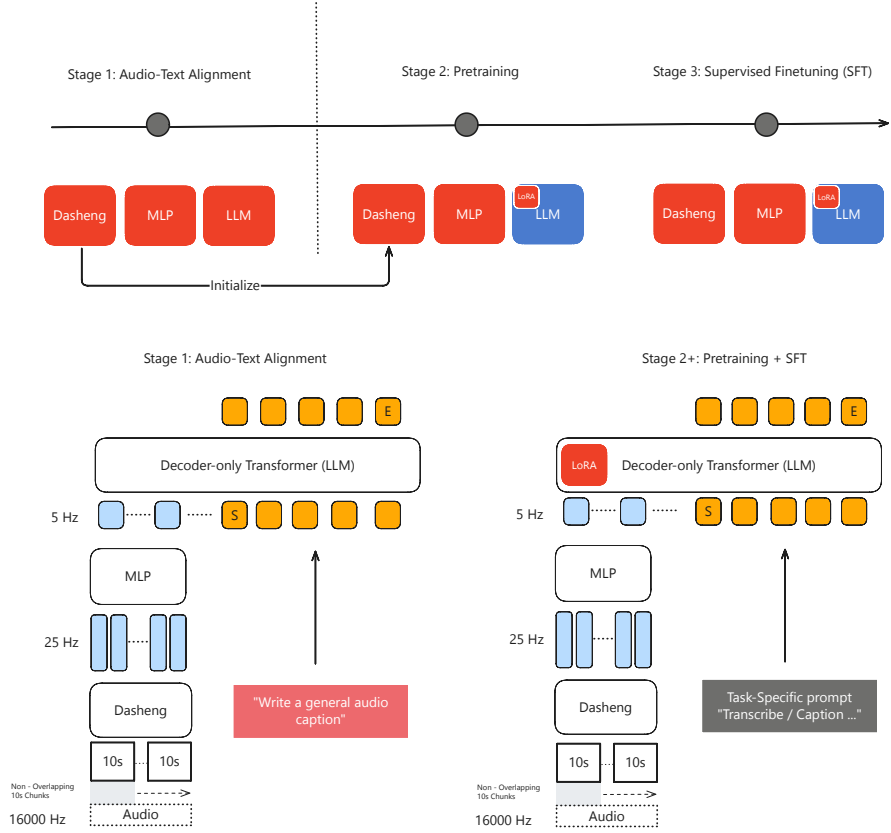}
    \caption{Proposed MiDashengLM framework. For all three stages, training is done with standard next-token prediction loss. Stage 1 aligns the audio encoder with the text modality, after which the audio encoder is taken and initialized for Stage 2.}
    \label{fig:framework}
\end{figure}

Our proposed framework can be seen in \Cref{fig:framework}. 
The framework is a common prefix-based large language model, where features of an audio encoder are mapped into the embedding space of an LLM via a multilayer perceptron (MLP) layer.
Our framework mainly differences from previous works in the following regards.

\paragraph*{Public data} Our approach only uses publicly available audio-text data for pretraining, supervised finetuning (SFT) and instruction tuning. 
All data sources are listed in \Cref{tab:speech_data,tab:env_data,tab:speech_meta_data,tab:music_data,tab:qa_data}.

\paragraph*{Audio-text alignment} 
Training LALMs is generally seen as an alignment problem, that aims to map audio features into a text-based space, such that an LLM can process these audio tokens.
In order to improve the training speed and performance, the vast majority of works utilize pretrained audio encoders.
One of the most prevalent pre-trained model is the Whisper encoder~\cite{radford2023robust}, as seen in models like LTU-AS~\cite{gong_ltuas}, Qwen-Audio~\cite{qwen_audio}, Qwen2-Audio~\cite{chu2024qwen2}, and Kimi-Audio~\cite{kimi_audio}, Mini-Omni~\cite{xie2024mini}, Llama-Omni~\cite{fang2024llamaomni}, R1-AQA~\cite{li2025reinforcement} and SALMONN~\cite{tang2023salmonn}. 
Other audio encoders such as HuBERT~\cite{hsu2021hubert}, HTS-AT~\cite{chen2022htsat}, AST~\cite{ast} and BEATs~\cite{chen2022beats} have also been utilized, often as secondary encoders to accommodate sound/music task knowledge.
To the best of our knowledge, this paper is the first to propose audio-text alignment via general captions, without relying on ASR or Sound event based models.
Further, we only utilize a \textit{single} general audio encoder that is jointly capable of processing speech, sound and music.

\paragraph*{Training efficiency} 
Even though transformer models are fully parallelize during training, they scale quadratically with regards to the input sequence length.
\begin{figure}[h]
    \centering
    \includegraphics[width=0.55\linewidth]{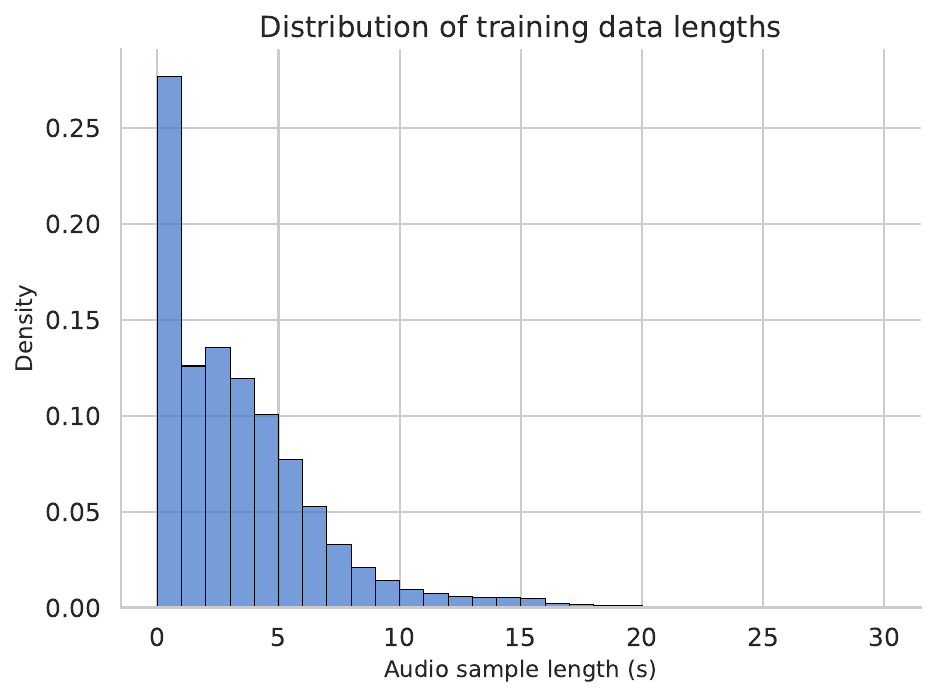}
    \caption{Histrogram plot of training data sample lengths.}
    \label{fig:data_length_distribution}
\end{figure}
Since most audio data used for LALM training has different lengths, one requires padding in order to batch samples towards a fixed sequence length.
One common way to significantly speedup training is by reducing the amount of padding by grouping samples with similar length together.
However, models such as Whisper natively does not support variable sequence lengths during training or inference and pads by default all inputs to a fixed duration of 30 seconds~\cite{chu2024qwen2}.
Changing this behavior can lead to significant performance degradation~\cite{jeffries2024moonshine,liu2025voxtral}.
We plot our training dataset's sample length distribution in~\Cref{fig:data_length_distribution}.
Since the majority of samples are between 1 and 10s long, padding to 30s would lead to inefficient training and inference, since the majority of encoder compute is wasted.
In contrast, our audio-encoder supports variable length inputs, significantly reducing the amount of padding and improve training efficiency.
More importantly, the majority of compute is done in the decoder, which benefits heavily from shorter sequences.
To further boost efficiency, we aggressively downsample the audio sequence length to a low framerate of 5 Hz, to accommodate fast training and inference speeds.

\section{Datasets}
\label{ssec:datasets}

MiDashengLM is trained solely on publicly available datasets during its pretraining and supervised finetuning phases.
All our training datasets are provided in \Cref{sec:data_sources}.
We further provide information about our novel general audio caption dataset.

\subsection{ACAVCaps and Multi-Expert Chain for Audio Tasks (MECAT)}
\label{ssec:acavcaps}

As discussed in \Cref{sec:motivation}, previous captioning datasets are insufficient mainly due to the lack of speech understanding and their monotonous data source mainly stemming from Audioset~\cite{gemmeke2017audio}, VGGSound~\cite{chen2020vggsound} and FSD50k~\cite{fonseca2021fsd50k}.
We identify that for our purposes, we would like a dataset that is publicly available and rich in content, containing multilingual speech, different types of music and a plethora of complex audio environments.
We identify ACAV100M~\cite{lee2021acav100m} as a plausible source dataset candidate for these purposes, since it has not been labeled for audio captioning before and contains little overlap with previously mentioned datasets.

Since ACAV100M lacks labels, we developed an efficient data curation pipeline.
We began by using CED-Base~\cite{dinkel2023ced} to predict AudioSet labels on a 2-second scale. 
We use this finer 2-second scale to enable our captions to capture temporal relationships.
Having obtained sound event labels, we further process the data using a plethora of different audio classification models, each tailored for a specific task.

\textbf{Speech Analysis:} This curation task identifies spoken language, distinguishes individual speakers, segments audio by speaker (diarization), detects speech emotion, classifies speaker gender and age and infers a transcript using Whisper~\cite{radford2023robust}.
\textbf{Vocal Analysis:} Beyond basic speech, this curation task refines vocal emotion detection, assesses vocal health, and analyzes unique vocal characteristics like pitch and timbre.
\textbf{Music Analysis:} For musical content, models classify music genre, recognize instruments, detect tempo, analyze music mood, and identify singing voices.
\textbf{Environmental Acoustics:} This part of the pipeline categorizes the acoustic scene, assesses audio quality, analyzes reverberation, and identifies various noise types.

\begin{figure}
    \centering
    \includegraphics[width=0.9\linewidth]{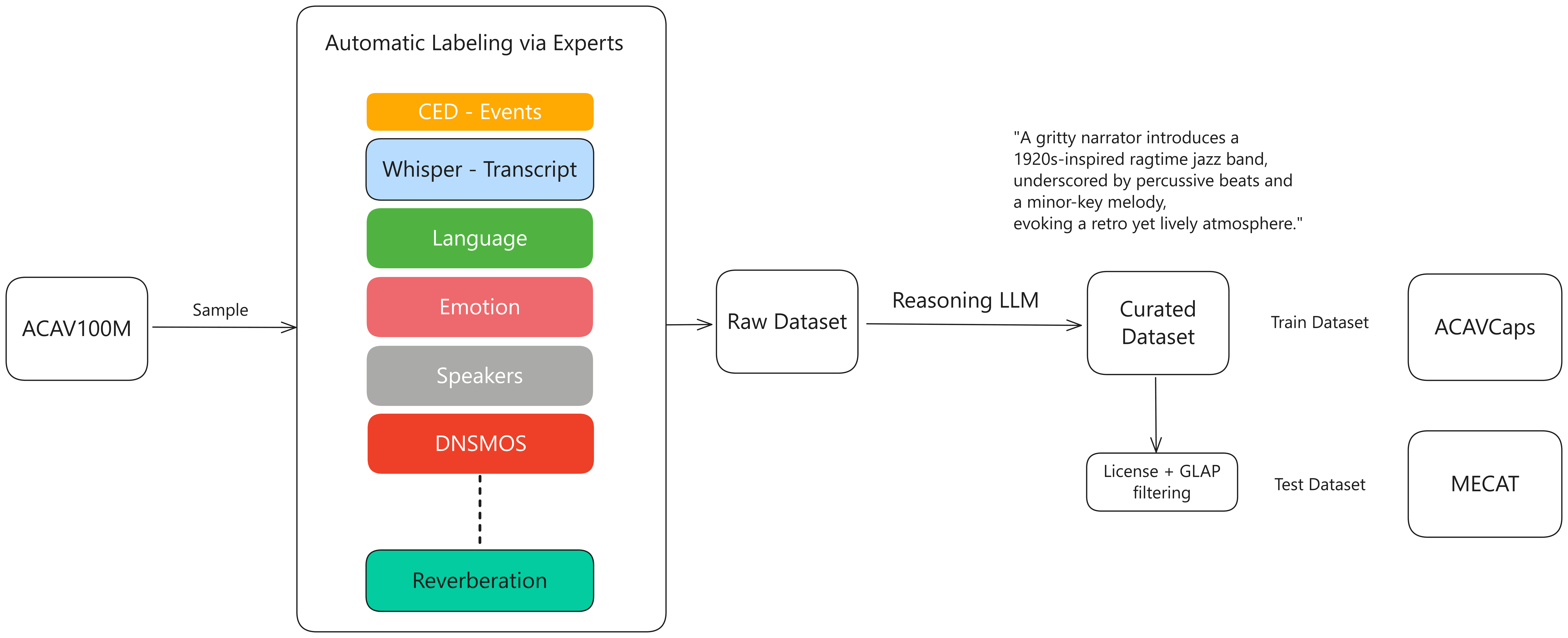}
    \caption{Our proposed data curation pipeline. We filter ACAV100M with an automatic pipeline, that predicts transcripts, sound events, sound quality and other meta information. A reasoning-LLM is then used to generate a caption from the provided meta information. The resulting curated dataset is then split into a training set (ACAVCaps) and a novel evaluation set Multi-Expert Chain for Audio Tasks (MECAT).}
    \label{fig:acavcaps_pipeline}
\end{figure}
\begin{table}[h]
\centering
\renewcommand{\arraystretch}{1.2}
\begin{tabular}{l p{0.8\linewidth}}
\toprule
{Category} & {Caption} \\
\midrule
Pure Speech & A female voice narrates a historical team competition (1966--1971) based on basketball rules, with intermittent synthetic speech modulation and variable acoustic reverberation. \\
Pure Sound & An outdoor scene with wind blowing, birds chirping, and a duck quacking, accompanied by significant background noise and low audio quality. \\
Pure Music & \textit{``If I were a zombie, I'd want your heart, not your brain''} --- A quirky electronic-pop anthem with gritty vocals, pulsing beats, and a dash of dark romance. \\
Mixed Music & The audio features a crowd cheering and clapping alongside electronic music with a synthesizer-driven, dark, and energetic soundscape. \\
Mixed Speech & A Russian voice demonstrates a synthesizer's capabilities over an experimental electronic backdrop, explaining its sound design and value in a gritty, vocal-fry tone. \\
Mixed Sound & A man speaks in English about entering a city and village, accompanied by the sounds of a running vehicle. \\
\bottomrule
\end{tabular}
\caption{A selection of our general audio captions generated by the proposed pipeline.}
\label{tab:samples_acavcaps}
\end{table}

Having obtained all these labels, we prompt a reasoning LLM (DeepSeek-R1~\cite{guo2025deepseek}) in order to generate a short audio caption.
The resulting curated audio caption dataset is then split into a train-set (ACAVCaps~\cite{niu2026acavcapsenablinglargescaletraining}) and test-set (\textbf{M}ulti-\textbf{E}xpert \textbf{C}onstructed Benchmark for Fine-Grained \textbf{A}udio Understanding \textbf{T}asks, MECAT). 
MECAT is extracted from the curated dataset by filtering each source video by license and finally performing GLAP~\cite{dinkel2025glap} to score the audio-text consistency.
A depiction of our pipeline can be seen in \Cref{fig:acavcaps_pipeline}.
MECAT will also be made publicly available~\cite{mecat}.
Lastly, we segment the dataset into six respective categories according to their CED labels, which can be seen in \Cref{tab:samples_acavcaps}.

Statistics about our resulting captioning training set can be seen in \Cref{tab:captions_compare}.
Notably, LAION-Audio-300M is a dataset that focuses on speech-only captions, neglecting sounds.
As we can see, our proposed dataset has a much richer vocabulary than previous approaches.
There are two main reasons for this.
First, since our captions summarize spoken content, the vocabulary naturally increases against other sound-event focused captions.
The second reason is the multilingual nature of our source dataset, where often transcripts from a foreign language are kept in the final caption e.g., ``A synthesized Spanish voice narrates a tense zombie confrontation: ``Repentinamente... golpe varias veces'' delivered with mechanical flatness amid variable reverberation and background noise.''

\begin{figure}
    \centering
    \includegraphics[width=0.65\linewidth]{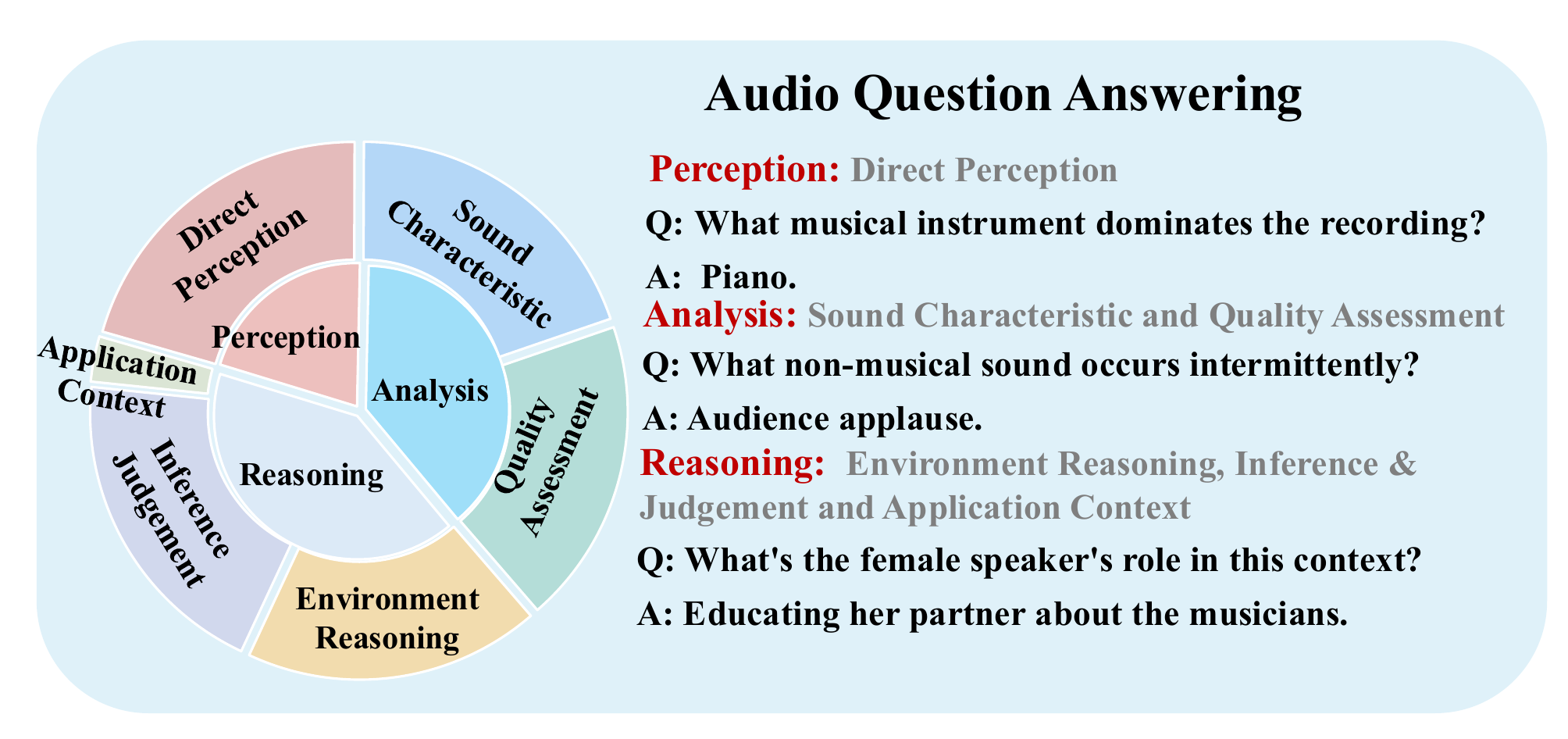}
    \caption{Subtasks of the proposed MECAT-QA testset.}
    \label{fig:mecat_qa}
\end{figure}

\paragraph*{MECAT-QA}
In MECAT-QA, each audio clip is paired with five question-answer pairs that span different categories and difficulty levels, resulting in over 100,000 total QA pairs. 
They are organized into three main cognitive categories: a) \textbf{Perception}, which consists of a single sub-category, \textit{Direct Perception}, focusing on the direct identification and naming of audio content and events. b)\textbf{Analysis}, which is composed of two sub-categories: \textit{Sound Characteristics}, for examining the acoustic properties of a sound (e.g., pitch), and \textit{Quality Assessment}, for evaluating the technical fidelity of the audio (e.g., noise level). c) \textbf{Reasoning}, which covers higher-level cognitive skills and is divided into three sub-categories: \textit{Environment Reasoning}, requiring the inference of the acoustic scene in which the sound occurs; \textit{Inference \& Judgement}, involving logical deductions and judgments based on the audio content; and \textit{Application Context}, testing the understanding of a sound's practical purpose or scenario.
A short introduction of available tasks and samples can be seen in \Cref{fig:mecat_qa}.

\begin{table}[htbp]
    \centering
    \caption{Comparison of publicly available captioning datasets. Datasets denoted with $^\ddagger$ contain multilingual captions. The number of unique words (\# Vocab) and the average sentence length are displayed.}
    \label{tab:captions_compare}
    \begin{tabular}{r|llll}
    \toprule
        Dataset & Labeling & \#Vocab & Avg. Sent & Source\\
        \midrule
        ClothoV2~\cite{drossos2020clotho} & \multirow{4}{*}{Manual} & 4366 & 11.32 & Freesound  \\
        AudioCaps~\cite{audiocaps} &  & 4844 & 8.70 & Audioset  \\
        MusicCaps~\cite{agostinelli2023music_caps} &  & 3730 & 47.17 & Audioset \\
        Songdescriber~\cite{manco2023songdescriber} &  & 1811 & 26.31 & MTG-Jamendo \\
        \midrule
        LPMusicCaps-MTT~\cite{doh2023lpmusiccaps} & \multirow{8}{*}{LLM} & 4045 & 25.04 & MagnaTagATune \\
        LPMusicCaps-MSD~\cite{doh2023lpmusiccaps} &  & 14049 & 37.06 & MillionSoundDatabase \\
        SoundVECaps~\cite{yuan2024soundvecaps} &  & 58401 & 31.48 & Audioset \\
        AutoACD~\cite{sun2024autoacd} &  & 20491 & 18.47 & Audioset \\
        AudiosetCaps~\cite{bai2024audiosetcaps} &  & 21783 & 28.13 & Audioset + VGGSound \\
        WavCaps~\cite{mei2023wavcaps} &  & 24592 & 7.84 & \makecell[l]{Audioset + BBC + \\ FreeSound + SoundBible } \\
        LAION-Audio-300M~\cite{laion300m} & & 451927 & 37.55 & ? \\
        \midrule
        Ours$^\ddagger$ & Reasoning-LLM & 644407 & 22.18 & ACAV100M \\
        
    \bottomrule
    \end{tabular}

\end{table}

\subsection{Training datasets and tasks}

\begin{figure}[htbp]
    \centering
    \includegraphics[width=1.0\linewidth]{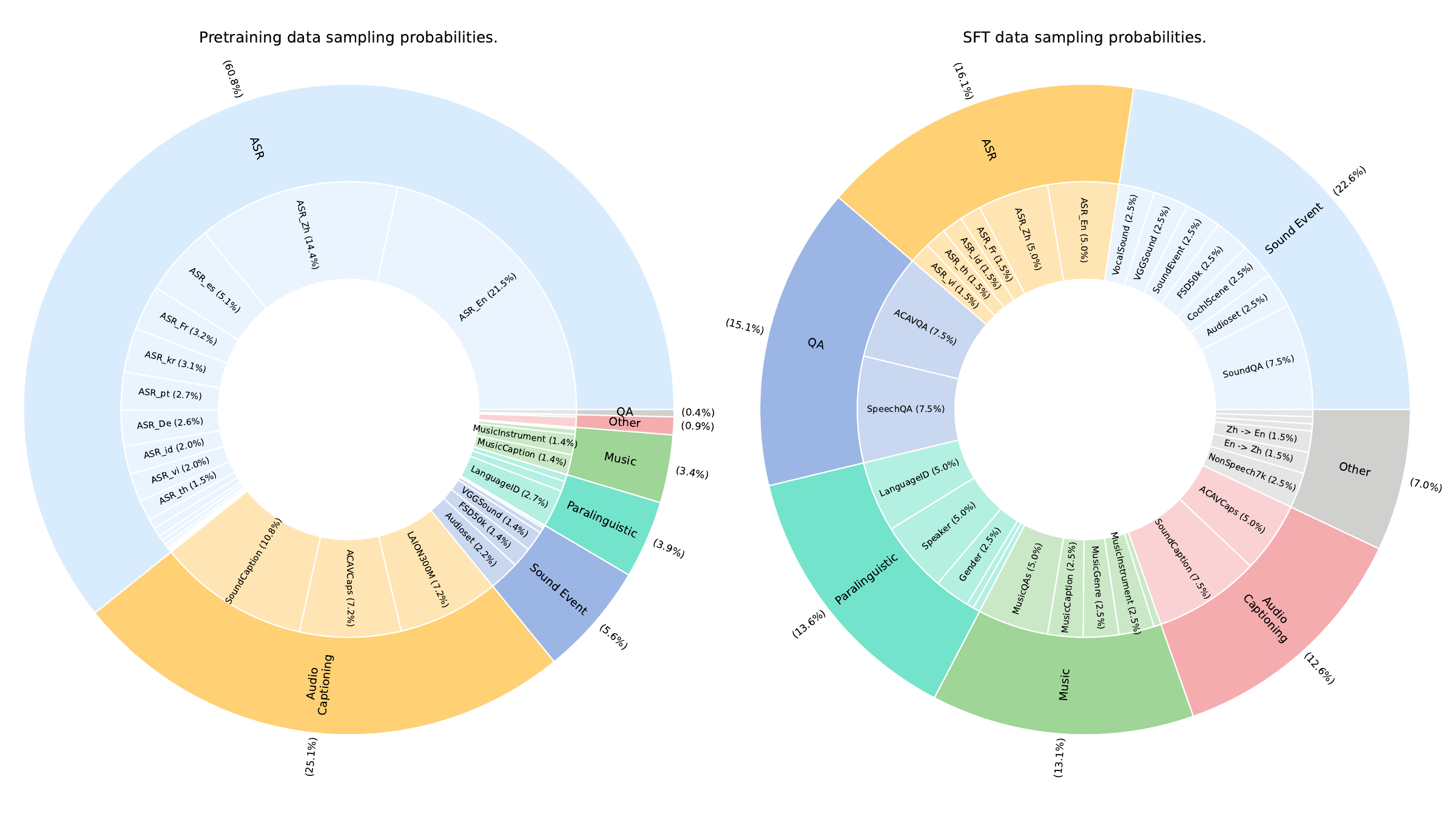}
    \caption{Pretraining and SFT sampling across datasets.}
    \label{fig:data_sampling}
\end{figure}

Our publicly available data sources, detailed in \Cref{sec:data_sources}, comprise approximately 1.1 million hours of data. 
Notably, approx. 90\% of the training data originates from public ASR datasets, while the remaining datasets are significantly smaller. 
If not properly treated, this would lead to inadequate performance for tasks other than ASR.
Data sampling can be viewed in \Cref{fig:data_sampling}. 
For audio-text alignment, we utilize the previously introduced ACAVCaps dataset (see \Cref{ssec:acavcaps}), which contains 38,000 hours of high-quality general captions.
We train for three epochs on ACAVCaps to align the audio encoder with text.
Following alignment, we pretrain MiDashengLM on the full 1.1 million hours of training data for approximately 1.4 epochs. 
After pretraining, we conduct supervised fine-tuning for one additional epoch on a curated subset of the pretraining data, totaling 352k hours. 
Further details on the datasets used can be found in \Cref{sec:data_sources}.

\section{Experimental Setup}

MiDashengLM is a standard Transformer-based encoder-decoder model~\cite{attention_is_all_you_need}, comprising a Transformer audio encoder and a text decoder.
The audio encoder builds upon Dasheng-0.6B~\cite{dinkel24b_interspeech_dasheng}, a \textit{frame-level} Vision Transformer (ViT)~\cite{Dosovitskiy_ViT} pretrained using the Masked Autoencoder (MAE) objective~\cite{he2022masked}, primarily on the ACAV100M dataset.
MiDashengLM exclusively supports 16 kHz audio inputs, and all input data is automatically resampled to this sampling rate.
Audio waveforms are converted into 64-dimensional mel-spectrograms, which Dasheng-0.6B processes by extracting 32 ms frame features with a 10 ms stride.

By default, Dasheng further downsamples the input features by a factor of four, producing high-level features at 40 ms intervals. 
As noted in \Cref{sec:motivation}, Dasheng supports variable-length inputs, with a maximum input length of 1008 frames (10.08 seconds). 
For longer inputs, we apply a non-overlapping sliding window approach, by forwarding each chunk through Dasheng, and concatenating the resulting frame-level features.
The complete hyperparameter configuration is documented in \Cref{tab:hyperparameters}, with a systematic comparison between our audio encoder and Whisper's architecture presented in \Cref{tab:whisper_vs_ours}. 
Training the full pipeline required roughly 19,200 GPU hours, or 10 days on 80 GPUs.


\begin{table}[h]
    \centering
    \caption{Audio encoder differences between our proposed model and the more common Whisper-Large v3.}
    \label{tab:whisper_vs_ours}
    \begin{tabular}{r|ll}
    \toprule
         & Whisper-Large v3 & Ours \\
        \midrule
        Parameters & 637.7M & 630.3M \\
        Pretraining data size & 5M & 270k \\
        Training Objective & ASR & General captions \\ 
        Context & 30s & 10s \\
        Known pretraining data? & \xmark & \cmark~\cite{lee2021acav100m}\\
        Open train code? & \xmark & \cmark~\cite{dinkel24b_interspeech_dasheng} \\
        Open weight? & \cmark & \cmark \\
        
        \bottomrule
    \end{tabular}

\end{table}

\paragraph*{Audio-text alignment} 
Pretraining for our Dasheng-based audio encoder is done via the masked-autoencoder (MAE) objective, which learns high-level audio features in a latent space.
However, a major difference between Whisper and our proposed Dasheng based encoder is that Whisper has been aligned with textual data (ASR).
Thus the first step of our MiDashengLM aligns the audio encoder with textual data.
For this alignment stage, we employ the ACAVCaps dataset, performing end-to-end fine-tuning of both the audio encoder and text decoder components.
Following alignment, we extract the trained audio encoder for initialization in subsequent pretraining and SFT phases.
During model development, we empirically evaluated two alternative approaches: (1) integration with a frozen large language model (LLM) and (2) low-rank adaptation (LoRA~\cite{hu2022lora}). 
However, both approaches yielded unsatisfactory audio encoder performance.
Audio-text alignment ran with an effective batch-size of 256 on 8 GPUs for one day.

\paragraph*{Text Decoder} 
The text decoder is initialized using Qwen2.5-Omni-7B~\cite{qwen25_omni}, a publicly available pretrained language model. 
For both pretraining and supervised fine-tuning phases, we employ LoRA to enhance parameter efficiency. 
The training objective minimizes the standard cross-entropy loss:
$$
\mathcal{L}_{ce} = -\log P(x_t | x_{1:t-1}, A),
$$
where $x_t$ is the current text token, $x_{1:t-1}$ represents the past text tokens, and $A$ denotes the audio features.

\paragraph*{Training}

All training procedures incorporate a linear learning rate warm-up spanning the initial 1,000 iterations, during which the learning rate increases from zero to the target value. 
Subsequently, the learning rate follows a cosine decay schedule, progressively decreasing to 10\% of its maximum value by training completion.
Notable differences between pretraining and SFT include: (1) a reduced learning rate during SFT, and (2) the expansion of trainable parameters influenced by LoRA. 
The training hyperparameters are provided in \Cref{tab:hyperparameters}.
Here ``all-linear'' modifies all projection layers within the decoder using LoRA, while ``q,v'' exclusively adapts the query and value matrices within the self-attention layers.

\begin{table}[h]
\centering
\caption{Decoder Hyper Parameters for MiDashengLM-7B and Training Configuration.}
\label{tab:hyperparameters}
\begin{tabular}{r|cc}
\toprule
& \multicolumn{2}{c}{Stage} \\
{Parameter} & Pretrain & SFT  \\
\midrule
Decoder-Size & \multicolumn{2}{c}{7B}  \\
Optimizer & \multicolumn{2}{c}{AdamW8bit}  \\
LoRA rank & \multicolumn{2}{c}{8}  \\
LoRA alpha & \multicolumn{2}{c}{32}  \\
LoRA dropout & \multicolumn{2}{c}{0.1}  \\
Audio-token framerate & \multicolumn{2}{c}{5 Hz} \\
\midrule
Learning rate & 1e-4 & 1e-5 \\
Weight decay & 0.01 & 0.1 \\
LoRA target & q,v & all-linear  \\
Batchsize & 10 & 8 \\
\bottomrule
\end{tabular}
\end{table}

\section{Results}
\label{sec:results}

We evaluate performance on each dataset's designated standard test/evaluation split.

\subsection{Audio encoder performance}
\label{ssec:audio_encder_performance}

To evaluate our audio-text alignment framework trained with general audio captions, we compare the resulting audio encoder against Whisper-Large V3. 
We employ the X-Ares benchmark~\cite{xares_zhang}, which evaluates frozen encoder embeddings through a lightweight MLP layer across three core audio domains: speech, music, and (environmental) sound.

\begin{table}[H]
\centering
\caption{Performance Comparison between our proposed captioning pretrained (Dasheng) model and Whisper-Large V3 (Whisper) using the X-Ares benchmark. For all metrics, higher is better and the best results are visualized in boldface.}
\label{tab:performance_encoder}
\begin{tabular}{rr|ccl}
\toprule
Domain & {Dataset} & {Ours} & {Whisper} & {Ours vs. Whisper} \\
\midrule
\multirow{11}{*}{Speech} & LibriCount & 61.9 & \textbf{64.4} & \color{xiaomibrightblue}{-3.9} \\
&LibriSpeech-100h & 85.4 & \textbf{90.0} & \color{xiaomibrightblue}{-5.1} \\
&LibriSpeech-MF & \textbf{98.5} & 94.9 & \color{xiaomicoral}{+3.8} \\
&VoxLingua33 & 92.3 & \textbf{97.4} & \color{xiaomibrightblue}{-5.2} \\
&Speech Commands V1 & 97.4 & \textbf{97.7} & \color{xiaomibrightblue}{-0.3} \\
&CREMA-D & \textbf{77.0} & 71.3 & \color{xiaomicoral}{+8.0} \\
&Fluent Speech Commands & \textbf{98.1} & 97.8 & \color{xiaomicoral}{+0.3} \\
&RAVDESS & \textbf{76.1} & 68.5 & \color{xiaomicoral}{+11.1} \\
&Vocal Imitation & \textbf{31.2} & 29.3 & \color{xiaomicoral}{+6.5} \\
&VocalSound & \textbf{93.2} & 91.5 & \color{xiaomicoral}{+1.9} \\
&VoxCeleb1 & \textbf{73.3} & 24.8 & \color{xiaomicoral}{+195.6} \\
\midrule
\multirow{7}{*}{Sound} & ASV2015 & \textbf{99.3} & 97.9 & \color{xiaomicoral}{+1.4} \\
&Clotho & \textbf{5.8} & 3.1 & \color{xiaomicoral}{+87.1} \\
&DESED & \textbf{53.7} & 22.6 & \color{xiaomicoral}{+137.6} \\
&ESC-50 & \textbf{94.3} & 62.5 & \color{xiaomicoral}{+50.9} \\
&FSD50k & \textbf{55.5} & 32.0 & \color{xiaomicoral}{+73.4} \\
&FSD18-Kaggle & \textbf{82.2} & 49.6 & \color{xiaomicoral}{+65.7} \\
&UrbanSound 8k & \textbf{87.9} & 75.7 & \color{xiaomicoral}{+16.1} \\
\midrule
\multirow{4}{*}{Music} &Free Music Archive Small & \textbf{67.2} & 58.9 & \color{xiaomicoral}{+14.1} \\
&GTZAN Genre & \textbf{88.6} & 71.8 & \color{xiaomicoral}{+23.4} \\
&MAESTRO & \textbf{54.5} & 0.0 & {$+\infty$} \\
&NSynth-Instruments & \textbf{72.2} & 63.5 & \color{xiaomicoral}{+13.7} \\
\bottomrule
\end{tabular}
\end{table}

As shown in \Cref{tab:performance_encoder}, our Dasheng-based encoder demonstrates strong performance across diverse audio classification tasks.
Comparative analysis reveals that while Whisper-Large v3 achieves superior results on 4 of 22 tasks, our encoder outperforms Whisper on the remaining 18 tasks.
Whisper outperforms our proposed encoder on tasks such as automatic speech recognition (ASR) by 5\% WER, speaker counting (LibriCount), spoken language recognition (VoxLingua33) and keyword spotting (Speech Commands V1).
All of those tasks are strictly speech-related.
On the other hand our proposed audio encoder outperforms Whisper-Large v3 on the majority of environment, music and sound classification tasks.
Largest gains are achieved for speaker recognition (VoxCeleb1, + 195\%), domestic sound event classification (DESED, + 137 \%) and Audio-text retrieval (Clotho, + 87\%).
These results demonstrate that audio-text alignment through general audio captions represents an effective approach for high-performance general-purpose audio understanding.

\subsection{Traditional dataset Benchmarks}

\begin{table}[h]
    \centering
    \caption{Comparison between the proposed MiDashengLM and baseline models.}
    \label{tab:model_comparison}
    \begin{tabular}{r|ccc}
    \toprule
        {Parameter} &\thead{MiDashengLM\\7B} & \thead{Qwen2.5-Omni\\7B} & \thead{Kimi-Audio-Instruct\\7B}  \\
        \midrule
        Encoder & Dasheng-based & Whisper-based & Whisper-based \\
        Decoder Parameters & {7B}  & 7B & 7B\\
        Audio-token framerate $\downarrow$ &  \textbf{5 Hz}  & 25 Hz & 12.5 Hz\\
        Audio-text alignment &  General caption  & ASR & ASR \\
        Capable of ASR ? & \cmark & \cmark & \cmark \\
        Known pretraining data ? & \cmark & \xmark & \xmark \\
        \bottomrule
    \end{tabular}
\end{table}

We evaluate our proposed MiDashengLM on common benchmarks against two strong baselines: Qwen2.5-Omni~\cite{qwen25_omni} and Kimi-Audio-Instruct~\cite{kimi_audio}.
Note that we exclusively compare with general audio understanding models that are capable of captioning as well as spoken language understanding in order to compare fairly, since there exist work solely optimized for captions only~\cite{ghosh2025audio,kong2024audio}.
A short overview about the models can be seen in \Cref{tab:model_comparison}.
For all subsequent results in tables and figures, we explicitly indicate decoder sizes using the following nomenclature: Qwen2.5-Omni-7B (Qwen2.5-Omni), Kimi-Audio-Instruct-7B (Kimi-Audio-Instruct) and MiDashengLM-7B (MiDashengLM).

\subsubsection{Audio captioning results}

Results for audio captioning can be seen in \Cref{tab:captioning_benchmark}, where we select FENSE~\cite{fense} as our primary audio caption metric.
For both music and audio (sound) captioning datasets, MiDashengLM outperforms consistently the baseline models.
The performance gains are particularly significant for general audio, with our model substantially outperforming baselines on AutoACD, while showing more modest improvements on music-specific benchmarks.

\begin{table}[htb]
\centering
\caption{Results for traditional music and audio captioning datasets. All results represent FENSE, where higher is better and best is in bold.}
\label{tab:captioning_benchmark}
\begin{tabular}{rr|ccc}
\toprule
Domain & Dataset &  \thead{MiDashengLM\\7B} & \thead{Qwen2.5-Omni\\7B} & \thead{Kimi-Audio-Instruct\\7B} \\
\midrule
\multirow{2}{*}{Music} 
 & MusicCaps    & \textbf{59.71} & 43.71 & 35.43 \\
 & Songdescriber & \textbf{45.39} & 45.31 & 44.63 \\
\midrule
\multirow{3}{*}{Sound} 
 & AudioCaps     & \textbf{62.18} & 60.79 & 49.00 \\
 & ClothoV2      & \textbf{49.20} & 47.55 & 48.01 \\
 & AutoACD       & \textbf{66.52} & 55.93 & 44.76 \\
\bottomrule
\end{tabular}

\end{table}

\subsubsection{MECAT}

Unlike traditional captioning datasets, MECAT provides a comprehensive evaluation framework across nine distinct domains: short captions, long captions, and pure/mixed categories of speech, sound, and music, along with environmental captions. 
This benchmark requires domain-specific caption generation—for instance, environmental captions must exclude spoken content, while pure-speech outputs should focus exclusively on verbal elements.
As shown in \Cref{tab:performance_mecat}, our results align with findings from standard audio captioning benchmarks (\Cref{tab:captioning_benchmark}).
From these results we observe that Kimi-Audio-Instruct performs poorly for captioning tasks.
Further, MiDashengLM, benefiting from its general captioning capabilities, surpassed the baselines by a significant margin.

\begin{table}[htb]
\centering
\caption{Model Performance Comparison on MECAT. All results represent FENSE, where higher is better and best is in bold.}
\label{tab:performance_mecat}
\begin{tabular}{r|ccc}
\toprule
{Task} &   \thead{MiDashengLM\\7B} & \thead{Qwen2.5-Omni\\7B} & \thead{Kimi-Audio-Instruct\\7B}  \\
\midrule
Content Long  & \textbf{60.11} & 48.34 & 40.83 \\
Content Short & \textbf{61.38} & 45.29 & 45.72 \\
\midrule
Pure Speech & \textbf{50.69} & 37.27 & 25.57 \\
Pure Sound & \textbf{53.78} & 46.60 & 35.75 \\
Pure Music & \textbf{66.17} & 50.68 & 39.54 \\
\midrule
Mixed Speech & \textbf{51.06} & 37.43 & 27.12 \\
Mixed Sound &\textbf{32.40} & 32.07 & 19.44 \\
Mixed Music & \textbf{59.50} & 34.71 & 16.18 \\
Environment & \textbf{51.38} & 47.84 & 16.66 \\
\midrule
Overall & \textbf{57.53} & 43.80 & 36.32 \\
\bottomrule
\end{tabular}
\end{table}

\subsubsection{Audio and paralinguistic classification}

We next evaluate our approach on paralinguistic tasks, with results detailed in \Cref{tab:paralingustic_benchmark}.
Note that we directly test the model's capabilities of each respective dataset, while other reports such as Kimi-Audio prompt the model with a choice of available labels.
For speaker verification (VoxCeleb1), we introduce a novel evaluation protocol that presents utterance pairs (same or different speakers) for binary classification.
We combine pairs of utterances - either from the same speaker or different speakers - and task the model with determining whether the two utterances originate from the same speaker or different speakers.
Performance across the ten tested tasks implicate that MiDashengLM outperforms baselines for speaker verification (VoxCeleb1), Language identification (VoxLingua107), Sound classification (VGGSound, FSD50k) and Music classification (NSynth, FMA).

\begin{table}[htb]
    \centering
    \caption{Results for audio classification and paralinguistic benchmarks. Best in bold.}
    \label{tab:paralingustic_benchmark}
    \begin{tabular}{rr|cccc}
\toprule
{Dataset} & {Metric} &  \thead{MiDashengLM\\7B} & \thead{Qwen2.5-Omni\\7B} & \thead{Kimi-Audio-Instruct\\7B}  \\ 
\midrule
VoxCeleb1 & \multirow{8}{*}{ACC $\uparrow$} & \textbf{92.36} & 59.71 & 82.72 \\
VoxLingua107 &  & \textbf{93.41} & 51.03 & 73.65 \\
VoxCeleb-Gender &  & 96.12 & 99.82 & \textbf{99.69} \\
VGGSound &  & \textbf{52.11} & 0.97 & 2.20 \\
Cochlscene &  & \textbf{74.06} & 23.88 & 18.34 \\
NSynth &    & \textbf{80.52} & 60.45 & 38.09 \\
FMA  &  & {63.73} & \textbf{66.77} & 27.91 \\
FSDKaggle2018 &   & \textbf{75.25} & 31.38 & 24.75 \\
\midrule
AudioSet & \multirow{2}{*}{mAP $\uparrow$} & \textbf{8.86} & 6.48 & 3.47 \\
FSD50K &  & \textbf{37.58} & 23.87 & 27.23 \\
\bottomrule
\end{tabular}

\end{table}

\subsubsection{Automatic speech recognition}

We assess ASR performance across all models using standard public benchmarks (see \Cref{tab:asr_benchmarks}). 
We would like to point out that audio-token framerate significantly impacts ASR performance, with higher rates improving performance at the expense of computational efficiency (\Cref{tab:model_comparison}).
These results align with our earlier findings in \Cref{tab:performance_encoder}, demonstrating that our encoder continues to trail the closed-source Whisper model - the audio encoder employed by both baseline systems.
Since MiDashengLM is a captioning model first and foremost, it's ASR performance suffers against the baselines on the traditional LibriSpeech dataset.
However, performance on larger test-sets such as People's Speech outperforms the Qwen2.5-Omni baseline.
Kimi-Audio performs best overall on English and Mandarin speech recognition, which is likely stemming from its large pretraining using English and Chinese ASR data.
However, MiDashengLM and Qwen2.5-Omni are both capable of ASR on different languages such as Indonesian, Vietnamese and Thai.
This suggests our encoder, despite no speech-specific training, develops surprisingly robust multilingual capabilities.

\begin{table}[htb]
    \centering
    \caption{Results for common ASR benchmarks. Results denoted with ``>100'' represent unsupported language, where the corresponding model only outputs English. All results represent WER/CER, where lower is better and the best result is displayed in bold.}
    \label{tab:asr_benchmarks}
    \begin{tabular}{rr|cccc}
\toprule
{Dataset} &  {Language} &  \thead{MiDashengLM\\7B} & \thead{Qwen2.5-Omni\\7B} & \thead{Kimi-Audio-Instruct\\7B}  \\

\midrule
LibriSpeech test-clean & \multirow{3}{*}{English} & 3.7 & 1.7 & \textbf{1.3} \\
LibriSpeech test-other & & 6.2 & 3.4 & \textbf{2.4} \\
People's Speech &  & 27.8 & 28.6 & \textbf{22.3} \\
\midrule
AISHELL2 Mic & \multirow{3}{*}{Chinese} & 3.2 & \textbf{2.5} & 2.7  \\
AISHELL2 iOS &  & 2.9  & \textbf{2.6}  &  \textbf{2.6}  \\
AISHELL2 Android &  & 3.1 & 2.7 &  \textbf{2.6} \\
\midrule
\multirow{3}{*}{GigaSpeech 2} & Indonesian & \textbf{20.8} & 21.2 & >100 \\
 & Thai & \textbf{36.9} & 53.8 & >100 \\
 & Viet & \textbf{18.1} & 18.6 & >100 \\

\bottomrule
\end{tabular}

\end{table}

\subsection{Question answering results}

Question answering (QA) performance results are presented in \Cref{tab:qa_results}. 
On closed QA benchmarks (MMAU~\cite{sakshi2024mmaumassivemultitaskaudio} and MuChoMusic~\cite{weck2024muchomusic}), MiDashengLM achieves superior performance with accuracies of 71.35\% and 66.30\%, respectively, outperforming all baseline models. 
This advantage extends to open QA tasks (MusicQA, AudioCaps-QA), where MiDashengLM maintains its leading position while Kimi-Audio-Instruct demonstrates the weakest performance, which is consistent with earlier captioning benchmark observations.

\begin{table}[htbp]
    \centering
    \caption{Results for question-answering datasets. For all results higher is better and best result are in bold.}
    \label{tab:qa_results}
    \begin{tabular}{rrr|ccc}
    \midrule
         {Dataset} & Subset &  {Metric} & \thead{MiDashengLM\\7B} & \thead{Qwen2.5-Omni\\7B} & \thead{Kimi-Audio-Instruct\\7B} \\  
         \midrule
         MuChoMusic~\cite{weck2024muchomusic} & &  ACC $\uparrow$ & \textbf{71.35} & 64.79 & 67.40 \\
         \midrule
         \multirow{4}{*}{MMAU~\cite{sakshi2024mmaumassivemultitaskaudio}} & Sound & \multirow{3}{*}{ACC $\uparrow$} & {68.47} & 67.87   & \textbf{74.17} \\
          & Music & & {66.77} &  \textbf{69.16} & 61.08 \\
          & Speech &  & \textbf{63.66} & {59.76} & 57.66 \\
          \cmidrule{2-6}
          & Average & & \textbf{66.30} & 65.60 & 64.30 \\  
         \midrule
         MusicQA~\cite{musicqa} & & \multirow{2}{*}{FENSE $\uparrow$} & \textbf{62.35} & 60.60 & 40.00 \\
         AudioCaps-QA~\cite{wang2024audiobench} & & & \textbf{54.31} & 53.28 & 47.34\\
        \bottomrule
    \end{tabular}
\end{table}

\subsubsection{MECAT-QA}

Lastly, we evaluate MiDashengLM on our proposed MECAT-QA dataset, a part of the publicly available MECAT benchmark~\cite{mecat}.
The dataset is a open QA dataset, which we evaluate using FENSE. 
As the results in \Cref{tab:mecat_qa_results} show, our proposed MiDashengLM outperforms the baselines by a significant margin on the MECAT-QA dataset.

\begin{table}[htbp]
\centering
    \caption{Results for MECAT-QA. Results represent FENSE, where higher is better and best result are in bold.}
    \label{tab:mecat_qa_results}
\begin{tabular}{r|ccc}
\toprule
Task & \thead{MiDashengLM\\7B} & \thead{Qwen2.5-Omni\\7B} & \thead{Kimi-Audio-Instruct\\7B} \\
\midrule
Direct Perception & \textbf{65.89} & 49.65 & 37.45 \\
Sound Characteristics &\textbf{62.10} & 43.81 & 32.48 \\
Quality Assessment & \textbf{61.76} & 40.47 & 19.24 \\
Environment Reasoning & \textbf{63.02} & 44.09 & 37.53 \\
Inference \& Judgement & \textbf{59.57} & 42.50 & 38.83 \\
Application Context & \textbf{60.12} & 41.92 & 33.82 \\
\midrule
Average & \textbf{62.08} & 43.74 & 33.22 \\
\bottomrule
\end{tabular}
\end{table}

\subsection{Inference speed}
\label{ssec:inference_speed}

\begin{figure}[h]
    \centering
    \includegraphics[width=0.9\linewidth]{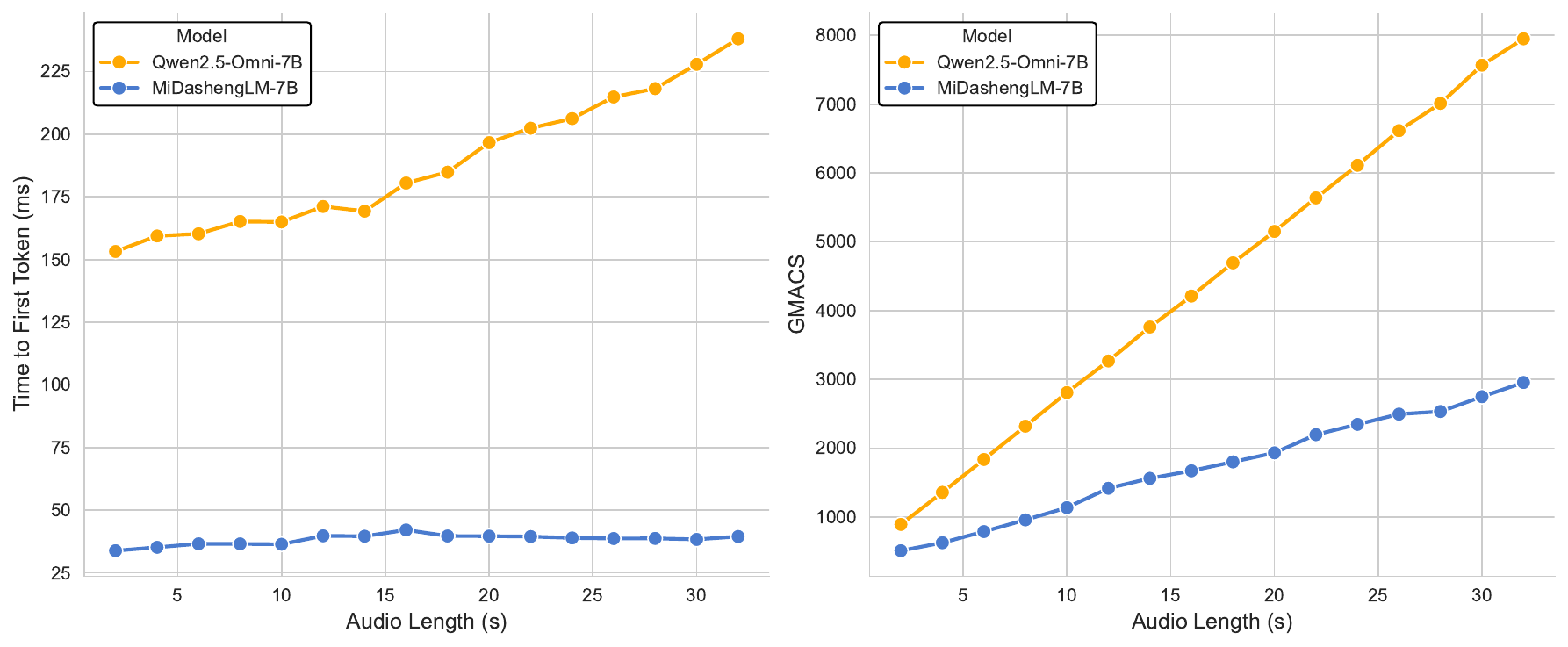}
    \caption{Time to first token (TTFT) and Giga Multiply-Add Operations per Second (GMACs) comparison between MiDashengLM-7B and Qwen2.5-Omni-7B.}
    \label{fig:TTFT_MACS_speed}
\end{figure}

A key advantage of MiDashengLM lies in its computational efficiency, encompassing both training speed (discussed in \Cref{sec:motivation}) and inference performance. 
In this experiments, we compare MiDashengLM with Qwen25-Omni-7B, as they utilizie the same text decoder backbone.
We provide results in regards to Time to first token (TTFT) latency and theoretical computation Giga Multiply-Add Operations per Second (GMACs), where results are displayed in \Cref{fig:TTFT_MACS_speed}.
As shown in \Cref{fig:TTFT_MACS_speed}, MiDashengLM achieves significantly lower TTFT than the baseline. 
We observe a speed improvement of up to 4$\times$ (160ms vs. 40ms) in regards to  TTFT.
Further throughput analysis in \Cref{tab:inference_speed} reveals a 3.2$\times$ speedup at comparable batch sizes and an overall potential speedup of 20.2$\times$ with larger batches.
These improvements stem from the better support for variable length inputs provided by Dasheng, as well as the optimized 5 Hz audio feature processing.

\begin{table}[h]
\centering
\caption{Throughput (samples/s) speed Comparison of MiDashengLM-7B and Qwen2.5-Omni-7B. Evaluation is done on a GPU with 80GB memory using bfloat16 for activations and parameters. All audio inputs are 30s long and output lengths are fixed to 100 tokens. OOM represents out of memory.}
\label{tab:inference_speed}
\begin{tabular}{r|cc|l}
\toprule
{Batch Size} &  \thead{MiDashengLM\\7B} & \thead{Qwen2.5-Omni\\7B} & {Speedup} \\
\midrule
1 & 0.65 & 0.45 & {{1.4$\times$}}  \\
4 & 2.42 & 1.21 & {2.0$\times$} \\
8 & 4.67 & 1.44 & {3.2$\times$} \\
\midrule
16 & 8.93 & \multirow{5}{*}{OOM} & 6.2$\times$  \\
32 & 14.36 &  & 10.0$\times$  \\
64 & 19.54 &  & 13.6$\times$ \\
128 & 24.26 &  & 16.8$\times$ \\
512 & 29.04 &  & 20.2$\times$  \\
\bottomrule
\end{tabular}
\end{table}

\section{Conclusion}

We present MiDashengLM, an efficient large audio language model (LALM) that advances the state of general audio understanding through several key innovations. 
First, we introduce a novel training paradigm using general audio captioning, enabled by our newly created ACAVCaps dataset and MECAT evaluation benchmark. 
This framework facilitates effective audio-text alignment, as demonstrated by our pretrained Dasheng-based encoder outperforming Whisper-Large V3 on 18 of 22 tasks in the X-Ares benchmark evaluation.
Notably, MiDashengLM achieves its strong performance while maintaining remarkable efficiency.
Trained exclusively on publicly available audio-text data, our model competes favorably against closed-source/closed-data alternatives (Qwen2.5-Omni and Kimi-Audio) across multiple domains including audio captioning, closed question answering, open question answering, sound event detection, and paralinguistic tasks. 
The model's computational advantages are particularly significant, delivering up to 20.2$\times$ faster inference speeds and up to 4 $\times$ reduced time-to-first-token latency compared to baseline approaches.




\clearpage
\bibliographystyle{IEEEtran} 
\bibliography{references} 



\newpage
\appendix

\section{Data sources}
\label{sec:data_sources}

\subsection{Speech datasets}
\label{ssec:speech_datasets}

\begin{table}[htbp]
\centering
\caption{Speech training data. The notation $^\dagger$ leverages Whisper to generate automatic transcripts by the authors. The column ``SFT ?'' indicates whether the dataset is used for supervised finetuning. By default all data is used for pretraining. }
\label{tab:speech_data}
\begin{tabular}{ll|rrr}
\toprule
{Data} & {Task} & {Length (h)} & SFT ? \\
\midrule
LibriSpeech~\cite{panayotov2015librispeech} & ASR & 960 & \cmark \\
LibriHeavy~\cite{kang2024libriheavy} & ASR & 50,000 & \xmark \\
GigaSpeech~\cite{chen2021gigaspeech} & ASR & 10,000 & \cmark \\
GigaSpeech 2~\cite{yang2024gigaspeech} & ASR & 30,000 & \cmark \\
WeNetSpeech~\cite{zhang2022wenetspeech} & ASR & 10,000 & \cmark \\
YODAS~\cite{li2023yodas} & ASR & 320,000 & \xmark \\
CommonVoice-17.0~\cite{ardila2019commonvoice} & ASR & 5,000 & \cmark \\
AISHELL-1~\cite{bu2017aishell1} & ASR & 100 & \cmark \\
AISHELL-2~\cite{du2018aishell2} & ASR & 1,000 & \cmark \\
AISHELL-3~\cite{AISHELL_3_2020} & ASR & 70 & \cmark \\
LJSpeech-1.1~\cite{ljspeech17} & ASR & 37 & \xmark \\
LibriTTS~\cite{zen2019libritts} & ASR & 585 & \xmark \\
MultiLingualSpokenWords~\cite{mazumder2021multilingual} & KWS & 5,000 & \xmark \\
Emilia~\cite{he2024emilia} & ASR & 101,000 & \cmark  \\
CovoST-v2~\cite{wang2021covost} & S2TT & 2,880 & \cmark \\
Fleurs~\cite{conneau2023fleurs} & S2TT & 1,224 &  \xmark \\
MSR-86K~\cite{li2024msr} & ASR, LangID & 86,000 & \cmark \\
ACAV100M-Speech$^\dagger$~\cite{lee2021acav100m} & ASR & 55,754 & \xmark \\
Must-C~\cite{di2019must} & ASR, S2TT & 1,000 &  \cmark \\
MLS~\cite{pratap2020mls} & ASR & 50,000 & \xmark \\ 
SpgiSpeech~\cite{o2021spgispeech} & ASR & 5,000 & \xmark \\
People's Speech~\cite{galvez2021peoplespeech} & ASR & 30,000 & \xmark \\
KeSpeech~\cite{tang2021kespeech} & ASR & 1,400 & \cmark \\
LAION-Audio-300M~\cite{laion300m} & Caption & 230,000 & \xmark \\
\midrule
\multicolumn{2}{l}{Total} & 997,010 & 258,410 \\
\bottomrule
\end{tabular}

\end{table}

\subsection{Sound and general audio datasets}

\begin{table}[H]
\centering
\caption{General Sound and Audio Datasets. ACAVCaps is utilized for audio-text alignment. The column ``SFT ?'' indicates whether the dataset is used for supervised finetuning. By default all data is used for pretraining.}
\label{tab:env_data}
\begin{tabular}{ll|rr}
\toprule
{Dataset} & {Task} & {Length (h)} & SFT ? \\
\midrule
FSD50k~\cite{fonseca2017freesound} & \multirow{7}{*}{Sound Event} & 77 & \cmark  \\
AudioSet~\cite{gemmeke2017audio} &  & 5,200 & \cmark \\
AudioSet-strong~\cite{audioset_strong} &  & 220 & \xmark \\
VGGSound~\cite{chen2020vggsound} &  & 540 & \cmark \\
FSDKaggle2018~\cite{fonseca2018generalfsd} &  & 20 & \cmark \\
FSDKaggle2019~\cite{fonseca2019audio} &  & 100 & \cmark \\
ARCA23k~\cite{arca23k} &  & 120 & \xmark \\
\midrule
AutoACD~\cite{sun2024autoacd} &  \multirow{7}{*}{Audio (Sound) Caption }& 5,200 & \cmark \\
AudioSetCaps~\cite{bai2024audiosetcaps} &  & 6,000 & \cmark \\
SoundVECaps~\cite{yuan2024soundvecaps} &  & 5,000 & \cmark \\
WavCaps~\cite{mei2023wavcaps} &  & 7,567 & \cmark \\
Audiocaps~\cite{audiocaps} &  & 100 & \cmark \\
Clothov2~\cite{drossos2020clotho} &  & 17 & \cmark \\
TACOS~\cite{primus2025tacos} &  & 98 & \cmark \\
\midrule
CochlScene~\cite{jeong2022cochlscene} & \multirow{2}{*}{SoundScape} & 500 & \cmark \\
BirdSet~\cite{birdset} &  & 7,000 & \xmark \\
\midrule
ACAVCaps & General Caption & 38,662 & \cmark \\
\midrule
\multicolumn{2}{l}{Total} & 76,421 & 69,081 \\
\bottomrule
\end{tabular}
\end{table}

\subsection{Speech and paralinguistic datasets}

\begin{table}[H]
\centering
\caption{Speech and sound paralinguistic datasets. The column ``SFT ?'' indicates whether the dataset is used for supervised finetuning. By default all data is used for pretraining.}
\label{tab:speech_meta_data}
\begin{tabular}{ll|rr}
\toprule
\textbf{Dataset} & \textbf{Task} & \textbf{Length (hours)} & SFT ? \\
\midrule
IEMOCAP~\cite{busso2008iemocap} & \multirow{7}{*}{Emotion} & 8 & \cmark \\
Meld~\cite{poria2018meld} &  & 12 & \cmark \\
SUBESCO~\cite{sultana2021subesco} &  & 9 & \xmark \\
RAVDESS-Speech~\cite{livingstone2018ravdess} &  & 2 & \xmark \\
RAVDESS-Song~\cite{livingstone2018ravdess} & & 1 & \xmark \\
CREMA-D~\cite{cao2014crema} &  & 4 & \xmark \\
ESD~\cite{zhou2021esd} &  & 29  & \xmark \\
\midrule
VocalSound~\cite{gong2022vocalsound} & \multirow{2}{*}{Vocal Sound classification} & 20 & \cmark\\
NonSpeech7k~\cite{rashid2023nonspeech7k} &  & 3 & \cmark \\
\midrule
VoxLingua107~\cite{valk2021voxlingua107} & \multirow{3}{*}{Language Identification} & 7,200 & \cmark \\
CommonLanguage~\cite{commonlanguage} &  & 45 & \cmark \\
YLACombe~\cite{ylacombe} &  & 5 & \xmark \\
\midrule
VoxCeleb1~\cite{nagrani17_interspeech} & Speaker verification & 76 & \cmark \\
CNCeleb~\cite{fan2020cnceleb} & Speaker verification & 2,100 & \cmark \\
VoxCeleb2~\cite{chung2018voxceleb2} & \makecell[l]{Speaker age\\Speaker verification\\Gender classification} & 1,000 & \cmark \\
VoxBlink1~\cite{lin2024voxblink} & Speaker verification & 1,300 & \cmark \\
VoxBlink2~\cite{lin2024voxblink2} & Speaker verification & 2,600 & \cmark \\
VoxTube~\cite{yakovlev2023voxtube} &  \makecell[l]{Speaker verification\\Language Identification\\Gender classification} & 5,200 & \cmark \\
LibriCount~\cite{libricount} & Speaker counting & 8 & \cmark \\
FluentSpeechCommands~\cite{lugosch2019speech} & Intent Classification & 17 & \xmark \\
speechocean762~\cite{zhang2021speechocean762} &  \makecell[l]{Gender\\Speaker age} & 5 & \xmark \\
ASVSpoof5~\cite{wang2024asvspoof} & Spoof detection & 603 & \xmark \\
\midrule
\multicolumn{2}{l}{{Total}} &  20,247 & 19,572 \\
\bottomrule
\end{tabular}
\end{table}

\subsection{Music Datasets}
\label{ssec:music_datasets}

\begin{table}[H]
\centering
\caption{Music-Related Datasets Overview. The column ``SFT ?'' indicates whether the dataset is used for supervised finetuning. By default all data is used for pretraining.}
\label{tab:music_data}
\begin{tabular}{ll|rr}
\toprule
{Dataset} & {Task} & {Length (h)} & SFT ? \\
\midrule
MusicCaps~\cite{agostinelli2023music_caps} &  \multirow{4}{*}{Music Caption} & 15  & \cmark\\
Songdescriber~\cite{manco2023songdescriber} & & 23 & \cmark \\
LPMusicCaps-MTT~\cite{doh2023lpmusiccaps} &  & 18 & \cmark \\
LPMusicCaps-MSD~\cite{doh2023lpmusiccaps} &  & 1,000 & \cmark \\
\midrule
VocalSet~\cite{wilkins2018vocalset} & Singing style identification & 10  & \xmark \\
FreeMusicArchive~\cite{nsynth2017} & Genre recognition & 610 & \cmark \\
MTG-Jamendo~\cite{bogdanov2019mtgjamendo} & \makecell[l]{Instrument classification\\Genre recognition} & 3,768 & \cmark \\
\midrule
NSynth~\cite{nsynth2017} & \multirow{4}{*}{Instrument classification} & 360 & \cmark \\
GoodSounds~\cite{bandiera2016good} &  & 28 & \cmark \\
chMusic~\cite{gong2021chmusic} &  & 1 & \cmark \\
CTIS~\cite{CTIS} &  & 1 & \cmark \\
\midrule
\multicolumn{2}{l}{{Total}} & {5,824} & 5,814 \\
\bottomrule
\end{tabular}
\end{table}

\subsection{Question Answering Datasets}
\label{ssec:qa_datasets}

\begin{table}[H]
\centering
\caption{Question answering datasets used in this work. Datasets denoted with $^\dagger$ have been modified from their original dataset by using an LLM to change captions into question-answer pairs. We display the number of questions and answers in each dataset as \# QA. The column ``SFT ?'' indicates whether the dataset is used for supervised finetuning. By default only AVQA, MusicQA ad ClothoAQA are used during pretraining.}
\label{tab:qa_data}
\begin{tabular}{ll|rr}
\toprule
{Dataset} & {Task} & \# QA & SFT ? \\
\midrule
AVQA~\cite{yang2022avqa} & \multirow{3}{*}{Environment QA} & 36,114 & \cmark \\
ClothoAQA~\cite{clotho_aqa} &  & 6175 & \cmark \\
TACOS$^\dagger$~\cite{primus2025tacos} &  &  40,019 & \cmark \\
\midrule
MusicQA~\cite{musicqa} & Music QA & 112,878 & \cmark \\
SIFT-50M~\cite{sift_50m} (closed) & Speech QA & 21,430,000 & \cmark \\
ACAV-QA$^\dagger$ & General QA & 24,371 & \cmark \\
\bottomrule
\end{tabular}
\end{table}

\section{Contributors}
Contributors are listed in Alphabetical order.
\begin{table}[htbp] 
    \begin{tabular}{l}
    {Heinrich Dinkel}\\
    {Gang Li}\\
    {Jizhong Liu}\\
    {Jian Luan}\\
    {Yadong Niu}\\
    {Xingwei Sun}\\
    {Tianzi Wang}\\    
    {Qiyang Xiao}\\
    {Junbo Zhang}\\
    {Jiahao Zhou}\\
    \end{tabular}
\end{table}

\end{document}